\def\flxdens{j(\rho)}
\def\rzero{\rho_{o}}
\def\nz{N_0}
\def\nt{N(t)}
\def\Mz{M(0)}
\def\Mt{M(t)}

\documentstyle[preprint,aps]{revtex}
\begin{document}
\draft

\title{Density Waves in Granular Flow: A Kinetic Wave Approach}
\author{Jysoo Lee and Michael Leibig}
\address{HLRZ-KFA J\"{u}lich, Postfach 1913, W-5170 J\"{u}lich, Germany}
\date{\today}
\maketitle

\begin{abstract} It was recently observed that sand flowing down a
vertical tube sometimes forms a traveling density pattern in which a
number of regions with high density are separated from each other by
regions of low density. In this work, we consider this behavior from
the point of view of kinetic wave theory. Similar density patterns are
found in molecular dynamic simulations of the system, and a well
defined relationship is observed between local flux and local density
-- a strong indicator of the presence of kinetic waves. The equations
of motion for this system are also presented, and they allow kinetic
wave solutions. Finally, the pattern formation process is investigated
using a simple model of interacting kinetic waves.
\end{abstract}


Systems of granular particles (e.g.\ sand) exhibit many interesting
phenomena, such as segregation under vibration or shear, density waves
in the outflow through tubes and hoppers, and probably most
strikingly, the formation of heap and convection cell under vibration
\cite{s84,c90,jn92,m92}. In granular flows through a narrow vertical
tube, P\"{o}schel found \cite{p92} that the particles do not flow
uniformly, but form high density regions which travel as coherent
structures with a velocity different from the center of mass
velocity. He also reproduced these density waves using molecular
dynamics (MD) simulations \cite{p92}.  However, the motion of these
high density regions and the mechanism which is responsible for their
formation are not fully understood.

In this paper, we present numerical and theoretical evidence that
these density waves are of a kinetic nature \cite{lw55}. Using MD
simulations, we measure the dependence of the particle flux on the
density. We find a well-defined flux-density relation -- an indication
that a kinetic wave theory describes the behavior. A direct
measurement of the velocity of these high density regions shows a
dependence on the mean density which is in good agreement with the
predictions from kinetic wave theory. On the theoretical side, we
consider one dimensional equations of motion for the density and the
velocity fields in the tube. These equations, together with Bagnold's
law for friction \cite{b54}, allow kinetic density wave solutions.

In order to understand the formation of these high density regions, we
consider the general problem of interacting kinetic waves. We first
show numerically that a system with an initially random density field
evolves to a configuration in which neighboring regions have a high
density contrast. At the early stage of development, we can show
analytically that the density contrast between nearby regions
increases linearly with time.

We first discuss the MD simulations of the system, and begin with a
brief description of the interparticle force laws that were used in
our calculations. The particles interact with each other (or with a
wall) only if they are in contact. The force that acts on particle $i$
due to particle $j$ can be divided into two components. The first,
$F^{n}_{j \to i}$, is parallel to the vector $\vec{r} \equiv \vec{R_i}
- \vec{R_j}$, where $\vec{R_i}$ and $\vec{R_j}$ are the coordinates of
the centers of particles $i$ and $j$ respectively. We refer to this as
the normal component. The second component, orthogonal to $\vec{r}$,
is the shear component $F^{s}_{j \to i}$. The normal component is
given by
\begin{mathletters}
\label{eq:f}
\begin{equation}
\label{eq:fn}
F^{n}_{j \to i} = k_n (a_i + a_j - |r|)^{3/2} - \gamma _{n} m_e
{\vec{v} \cdot \vec{r} \over |r|},
\end{equation}
where $a_i (a_j)$ is the radius, and $m_i$ $(m_j)$ the mass of
particle $i$ $(j)$. Also, $m_e$ is the effective mass $m_i m_j / (m_i
+ m_j)$, and $\vec{v} \equiv d\vec{r}/dt$. The first term in Eq.\
(\ref{eq:fn}) is the Hertzian elastic force, where $k_{n}$ is a
material dependent elastic constant. The second term is a velocity
dependent friction term, where $\gamma _{n}$ is a normal damping
coefficient. The shear component is given as
\begin{equation}
\label{eq:fs}
F^{s}_{j \to i} = -\gamma _{s} m_e {\vec{v} \cdot \vec{s} \over |s|},
\end{equation}
\end{mathletters}
where $\vec{s}$ is defined by rotating $\vec{r}$ clockwise by $\pi/2$.
The shear force, Eq.\ (\ref{eq:fs}), is simply a velocity dependent
friction term similar to the second term in the normal component.
Finally, we must specify the interaction between a particle and a
wall. The force on particle $i$, in contact with a wall, is given by
Eqs.\ (\ref{eq:f}) with $a_{j} = \infty$ and $m_{j} = \infty$. The
choice of the interactions defined by Eqs.\ (\ref{eq:f}) is rather
typical in the MD simulations of granular material
\cite{mdgranule}. A detailed explanation of the interaction is given
elsewhere \cite{lh93}.

For simplicity, we study granular flows in $2$ dimensions and use a
fifth order predictor-corrector scheme to integrate the equations of
motion, calculating both the positions and velocity of each particle
at all times. The tube is modeled by two vertical sidewalls of length
$L$ with a separation $W$, and we apply a periodic boundary condition
in the vertical direction. Between the sidewalls, particles of radii
$0.1$ are initially filled with a uniform density of $\rho _o$
(throughout this paper, numerical values are given in CGS unit). The
particles begin to move under the influence of gravity, and soon reach
a steady state, where the gravitational force is balanced by the
frictional force from the interactions with the sidewalls.

In Fig.\ \ref{fig:mdtube}, we show the time evolution of the density
and the velocity fields for $L = 15$ and $W = 1$, measured at every $5$
ms. At a given time, we divide the tube into $15$ vertical regions of
equal length, and measure the density and the average velocity in each
region. These fields are displayed as a vertical strip of square
boxes, where each box corresponds to a region in the tube. The
grayscale of the box is proportional to the value of the field in that
region. The parameters we used in this simulation are $k_{n} = 1.0
\times 10^{6}, \gamma _{n} = \gamma _{s} = 5.0 \times 10^{2}$, with
the time step $5.0 \times 10^{-5}$. The initial density $\rho _{o}$ is
25 particles per unit area. In the figure, we find (1) a region of
large density fluctuations is formed out of the initially uniform
system, (2) the fluctuations seem to travel with almost constant
velocity (different from the center of mass velocity), and (3) there
seems to be strong correlation between the density and the velocity
fields. These findings remain true for the simulations we have
performed with different values of $\gamma$, $k_{n}$ and $\rho _{o}$,
except when $\rho _o$ is very small, where a steady state is not
reached. These traveling density patterns were first observed in the
simulations by P\"{o}schel \cite{p92}.

In order to quantitatively study the correlation between the density
and the velocity fields, we measure the local particle flux as a
function of the local density in the following manner. Once the system
has reached a steady state, we measure the mean velocity $v_i$ and the
density $\rho _i$ in region $i$. The flux $j(\rho )$ for a given
density $\rho$ is then taken to be $\rho \cdot \langle v (\rho)
\rangle$, where $\langle\rangle$ is a time average over all regions
which had a particular density $\rho$. The flux-density curve,
obtained by averaging over $10{,}000$ iterations, are shown in Fig.\
\ref{fig:jrho}.  Here, the parameters are the same as those of Fig.\
\ref{fig:mdtube}. The fact that a well-defined flux-density curve
exists suggest that the density waves (traveling density
fluctuations) are kinetic in nature.  Furthermore, the flux-density
curve for the granular flow resembles that of a traffic flow, which is
considered as a prime example of the systems which shows kinetic waves
\cite{lw55}.

One additional piece of evidence that the density waves are of a
kinetic nature is their dependence on the initial density $\rho _{o}$.
The theory of kinetic waves predicts \cite{lw55} that small density
fluctuations in a uniform density background $\rho _{o}$ travel with a
velocity
\begin{equation}
\label{eq:kvel}
U(\rho _{o}) = {dj(\rho ) \over d\rho } \mid _{\rho = \rho _{o}},
\end{equation}
which is the slope of the flux-density curve at the mean density. We
thus expect a large negative velocity for small $\rho_{o}$, a decrease
to zero velocity at $\rho_{o} \approx 15$, with an increasingly large
positive velocity as $\rho_{o}$ is increased further. To check this,
we measure the wave velocities for several values of $\rho _{o}$
(keeping all other parameters fixed as above). Writing the mean
density $\rho _o$ and the corresponding velocity $U(\rho _o)$ as
$(\rho _o, U(\rho _o))$, we find $(10.0, -41 \pm 2),\ (15.0, 5 \pm 9),\
(18.7, 12 \pm 11)$ and $(22.5, 113.0 \pm 4)$, which are all consistent with
the above prediction.

We now consider the theoretical aspect of the density waves. Consider
the equations of motion which govern the time evolution of the
density $\rho (x,t)$ and the velocity $v(x,t)$ fields for a granular
flow in a vertical tube. The first equation is that of mass
conservation
\begin{mathletters}
\label{eq:motion}
\begin{equation}
{\partial \over \partial t}\rho + {\partial \over \partial x} (\rho v)
= 0,
\end{equation}
and the second is a momentum conservation equation
\begin{equation}
\rho {\partial \over \partial t}v + \rho v {\partial \over \partial x}
v = F(x,t),
\end{equation}
\end{mathletters}
where $F(x,t)dx$ is the total amount of force acting on the particles
in a region $[x,x+dx]$. The force $F(x,t)$ has three
contributions---gravity, internal pressure, and friction from the
sidewalls. The exact form of the internal pressure and the friction is
not known. Here, we use Bagnold's law, which is believed to be valid
in the grain inertia regime \cite{b54}. Therefore, the force $F(x,t)$
is
\begin{equation}
\label{eq:bagnold}
F(x,t) = -\rho g W - \text{sign}(v) \rho _{B} f_{xy}(p) v^{2}- D
{\partial \over \partial x}[\rho _{B} f_{xx}(p) v^{2}].
\end{equation}
Here, $g$ is gravitational acceleration, $\rho _{B}$ the density of
the material which forms the particles, $p$ the packing fraction
($\rho = \rho _{B} p$) and $D$ is the diameter of the particles. We
assume the thickness of the shear layer to be of order of $D$. Also,
$f_{xx}$ and $f_{xy}$ are geometry dependent functions, which contain
the information about the density dependence of the forces.

The uniform density solution of Eq.\ (\ref{eq:motion}) with the force
given by Eq.\ (\ref{eq:bagnold}) is
\begin{eqnarray}
\label{eq:uniform}
\rho (x,t) = & \rho _{B} p_{o} \cr
v (x,t) = & - \sqrt{p_{o}gW / f_{xy}(p_{o})}
\end{eqnarray}
If we add small density fluctuation $p = p_{o} + dp$ in the uniform
density flow, it can be shown \cite{l93b} that the fluctuation travels with
a velocity
\begin{equation}
\label{eq:kvel2}
U(p_{o}) \simeq - {1 \over 2} \sqrt{p_{o}gW / f_{xy}(p_{o})} \cdot {3
f_{xy}(p_{o}) - p_{o}df_{xy}(p_{o})/dp \over f_{xy}(p_{o})}.
\end{equation}
Equations (\ref{eq:uniform}) and (\ref{eq:kvel2}) are exactly what one
expects if the kinetic wave theory is to apply -- uniform flow is a
solution to the equations of motion, and density fluctuations travel
with a density dependent velocity.

Thus, it is clear then that the motion of the density pattern can be
understood by applying the ideas of kinetic wave theory. However, this
basic formalism only describes the motion of a pre-existing density
pattern. It does not explain the observation that regions with large
density contrasts are being formed out of the uniform background. Our
simulations show that the large scale density pattern begins as a
collection of small fluctuations in the density. These small
fluctuations grow in time and a pattern emerges in which large
density contrasts exist between neighboring regions. The evolution to
such a state can be understood by considering the system as set of
interacting kinetic waves. A detailed treatment of the general problem
of interacting kinetic waves can be found elsewhere \cite{l93}; in
this paper, we present only the results from a simple model for the
pattern evolution process in sand flowing down a tube.

Consider the early stages of the flow in which the density of sand is
nearly uniform at $\rho \approx \rzero$. Because of the roughness of
the grains, the roughness of the walls or from the stochastic nature
of the inelastic collisions, small density fluctuations appear in the
system. In the interacting density wave approach, we treat the
fluctuations as a set of distinct density regions with interfaces
whose velocities are determined by a discrete form of Eq.\
(\ref{eq:kvel}). In this case, the interface separating a region of
density $\rho_1$ from a region with density $\rho_2$ moves with a
velocity, $U(1,2)$, given by
\label{eq:discrete}
\begin{equation}
U(1,2) = {j(\rho_1) - j(\rho_2) \over \rho_1 - \rho_2},
\end{equation}
which is the kinetic wave theory result for interfacial velocities
involving finite density differences \cite{lw55}.  The evolution of
the system is determined by the motion of the interfaces and, as shall
be shown, the nature of their interactions leads to a final state in
which large density contrasts occur.

In the computational and analytic results that follow, we choose a
specific form for the density fluctuations in the system. In our
model, the initial positions of the interfaces are taken as a set of
$\nz$ points placed randomly on the interval $[0, L]$, with regions
between successive interfaces being assigned a density randomly in the
range $[\rzero - W, \rzero + W]$. The principal virtues of this model
are its simplicity and the fact that there are no correlations in the
initial state which might influence the final structure. A more
realistic model for the fluctuations of the system would require a
microscopic understanding of each specific source of noise.

It is also necessary to choose a form for the flux curve $\flxdens$.
We have taken the parabolic form
\begin{equation}
\flxdens = J_o{\rho\over R} \left(1 - {\rho\over R}\right),
\end{equation}
where $R$ is the density at which no flow occurs, and $J_o$ is one
quarter of the maximum flux of the system. This curve was chosen for
several reason. The first is that its simplicity eases some of the
hardships of analytical calculations. The second reason is that for
density fluctuations over a sufficiently small range, the true flux
response can be approximated by this form (with $R$ and $J_o$ being
fitting parameters). And finally, it is a first approximation to the
form observed for the $\flxdens$ observed in Fig.\ \ref{fig:jrho}.

Numerical simulation of this system is a very straight-forward
exercise. The values of the densities in two adjacent regions
determine the velocity of the associated interface. Consider three
successive density regions A, B and C. During the course of the
simulation, the interface A-B may encounter the interface B-C. This
indicates that all of the mass that was inside region B has been
completely ``swallowed up'' by the regions A and C. In this case, the
interfaces A-B and B-C are replaced by a single A-C interface. The
velocity of this interface can be calculated from the densities in
regions A and C. Thus, it is a matter of tracking all of the
interfaces, checking for collisions, and when they occur, replacing
the two old interface with a single new one. Therefore, this technique
does not allow for any density values other than those initially
present, and the number of interfaces is always decreasing. For
convenience, the simulations were done using periodic boundary
condition.

The first set of results shown below are from a simulation in which
there are 400 interfaces initially placed randomly in the interval
$[0, 1]$ (i.e.\ $L = 1$).  We also choose the values $J_o = 1$ and $R
= 1$.  The densities are chosen at random from the interval $[0.3,
0.8]$ (i.e.\ $\rzero = .55R$, $W = .25R$). Figure
\ref{fig:kevol}(a) shows the initial density configuration, while
figure \ref{fig:kevol}(b) shows the system after a time $t = .486$
(where time is measured in the units of $RL/J_o$), and there are only
33 interfaces which remain along the interval. The system has evolved
to a state in which the density contrast is very high between
neighboring regions, and this behavior was observed for all values of
$\rho_o$ and $W$.

This increase in the density contrast can be characterized
quantitatively in the following way. Let the density of each region be
$\rho_i$, with $i$ indexing the different regions, and $N(t)$ be the
number of regions at time $t$. Define the quantity
\begin{equation}
\Mt \equiv {1\over \nt} \sum_{i = 1}^{\nt} |\rho_i - \rho_{i + 1}|,
\end{equation}
where $\rho_{N(t)+1} \equiv \rho_1$. The larger the value of $\Mt$ the
larger the density contrast between neighboring regions. Figure
\ref{fig:mt} shows the quantity $\Mt - \Mz$ averaged over 10
simulations with $\nz = 10{,}000$ interfaces. At early times, there is
a linear increase in $\Mt$ with a crossover to a nearly constant value
at late times.

At early times, it is possible to calculate $\Mt$ analytically and the
results are shown as the dotted line in Figure \ref{fig:mt}. In this
regime, the changes in $\Mt$ are dominated by the interaction of
interfaces whose movements are determined by the initial configuration
of the system.  The calculation averages over all possible
configurations of the initial random densities and interfaces,
determines the time at which each interface collision occurs and how
much that collision changes the value of $\Mt$. In this regime, the
agreement with the simulation is good. It is also possible to show
exactly that $\Mz = 2W/3$.  At later times, after there have been many
collisions between interfaces, the structure of the system depends on
the nature of the earlier evolution. Thus, this long time behavior is
much more difficult to calculate. The results from the calculation
described above break down in this regime because the distribution of
density regions is no longer that of the initial random distribution.

At long times $\Mt \approx 2W$. Thus, the density contrast at long
times is, on average, as large as the largest density contrasts
present in the initial configuration. It turns out that the
interacting kinetic waves do not create large contrasts. Rather, the
interfaces from the initial distribution which survive are those that
have a very large density contrast \cite{l93}. Thus, while the noise
in the system may provide a variety of such contrasts, the interacting
kinetic waves will destroy all but the very largest.

This paper outlines a kinetic wave approach to understanding the
density patterns observed in sand flow along a vertical tube (many of
the details omitted here can be found in references \cite{l93b} and
\cite{l93}). However, these ideas certainly do not constitute a
complete theory for the patterns observed in the experimental system.
The role that the flow of air plays in this process \cite{air}, as
well as the sources of noise in the system, are certainly not well
understood. Further experimental investigation of these issues would
be most enlightening.  From a theoretical point of view, it is not
clear whether the frictional force at the wall and the internal
pressure obey Bagnold's law. While this form has been observed in the
sheer cell geometry \cite{b54,shear}, there has been no direct
measurement of the frictional force for gravity driven flow. Finally,
it is known that the interface between two regions of differing
densities may not be a stable structure \cite{lw55}, and that
diffusive effects may strongly influence the long time behavior of a
system of interacting kinetic waves.

\bigskip

The authors would like to thank the members of the HLRZ Many Body
Group for stimulating discussion throughout this work.

\begin{figure}
\caption{Time evolution of (a) density and (b) velocity fields.
These simulations were done with a tube of width $W=1$ and length
$L=15$. Fields at a given time are shown as a vertical line of small
boxes. The grayscale of each box is proportional to the value of the
density or velocity in that region of the tube. Regions of high
density are formed, and travel with almost constant velocity.}
\label{fig:mdtube}
\end{figure}

\begin{figure}
\caption{Local flux as a function of local particle density. This
curve was found for a tube with width $W=1$ and length $L=15$,
obtained by time averaging. The parabolic shaped curve resembles the
flux-density relation in traffic flows.}
\label{fig:jrho}
\end{figure}

\begin{figure}
\caption{Evolution of interacting kinetic waves. Both strips
use a linear grayscale with white indicating $\rho_o = 0$ and black
$\rho_o = R$, the jamming density. (a) Shows the initial configuration
of 400 interfaces with densities in the range $[0.3R, 0.8R]$. (b)
Shows the configuration when only 33 interfaces remain, and
illustrates the tendency for alternating high and low density
regions.}
\label{fig:kevol}
\end{figure}

\begin{figure}
\caption{Contrast, $M(t)$, as function of time, $t$. The solid line
shows the results from averaging over 10 simulations with $N(0) =
10{,}000$, and densities chosen in the range $[0.3, 0.8]$. At early
times, the increase in contrast is linear and at long times it becomes
a constant.  The dashed line shows the results from an analytical
calculation of the short time behavior.}
\label{fig:mt}
\end{figure}


\begin{references}
\bibitem{s84} S. B. Savage, Adv. Appl. Mech. {\bf 24}, 289 (1984); S. B.
Savage, {\it Disorder and Granular Media} ed. D.  Bideau,
North-Holland, Amsterdam (1992).

\bibitem{c90} C. S. Campbell, Annu. Rev. Fluid Mech. {\bf 22}, 57 (1990).

\bibitem{jn92} H. M. Jaeger and S. R. Nagel, Science {\bf 255}, 1523
(1992).

\bibitem{m92} A. Mehta, Physica A {\bf 186}, 121 (1992).

\bibitem{p92} T. P\"{o}schel, HLRZ preprint 89/92 (1992).

\bibitem{lw55} M. J. Lighthill and G. B. Whitham, Proc. Roy. Soc. A
{\bf 229}, 281 and 317 (1955).

\bibitem{b54} R. A. Bagnold, Proc. R. Soc. London A {\bf 225}, 49
(1954).

\bibitem{mdgranule} For example, P. A. Cundall and O. D. L. Strack,
G\'{e}otechnique {\bf 29}, 47 (1979); P. K. Haff and B. T. Werner,
Powder Technol. {\bf 48}, 239 (1986); Y. M. Bashir and J. D. Goddard,
J. Rheol. {\bf 35}, 849 (1991); P. A. Thompson and G. S. Grest, Phys.
Rev. Lett. {\bf 67}, 1751 (1991); G. Ristow, J. Physique I {\bf 2},
649 (1992); Y-h. Taguchi, Phys. Rev. Lett. {\bf 69}, 1371 (1992); J.
A. C. Gallas, H. J. Herrmann and S. Soko\l owski, Phys. Rev. Lett.
{\bf 69}, 1375 (1992); D. C. Hong and J. A. McLennan, Physica A {\bf
187}, 159 (1992).

\bibitem{lh93} J. Lee and H. J. Herrmann, J. Phys. A {\bf 26}, 373
(1993).

\bibitem{l93b} J. Lee, HLRZ preprint 44/93 (1993).

\bibitem{l93} M. Leibig, HLRZ preprint 42/93 (1993).

\bibitem{air} D. Bideau, private communication; A. Hansen, private
communication.

\bibitem{shear} For example, S. Savage and M. Sayed, J. Fluid Mech.
{\bf 142}, 391 (1984); C. Campbell, J. Fluid Mech. {\bf 203}, 449
(1989).

\end{references}
\end{document}